\documentclass[aps,pra,preprint,groupedaddress,showpacs]{revtex4}%
\usepackage[latin1]{inputenc}
\usepackage{eurosym}
\usepackage{makeidx}
\usepackage{graphicx}
\usepackage{amsmath}
\usepackage{amsfonts}
\usepackage{fancyhdr}
\usepackage{setspace}
\usepackage{amssymb}
\usepackage{anysize}%
\providecommand{\U}[1]{\protect\rule{.1in}{.1in}}
\begin{document}
\title{Exact density oscillations in the Tonks-Girardeau gas and their optical detection}
\author{Mihály G. Benedict} \email{benedict@physx.u-szeged.hu} \author{Csaba Benedek} \author{Attila Czirják}
\affiliation{Department of Theoretical Physics, University of Szeged, 
\\
H-6720 Szeged, Tisza Lajos körút 84-86, Hungary}

\begin{abstract}
We construct the exact time dependent density profile for a superposition of
the ground and singly excited states of a harmonically trapped one dimensional
Bose-Einstein condensate in the limit of strongly interacting particles, the
Tonks-Girardeau gas. Results of an off resonant light scattering experiment
probing the system could allow to determine the number of particles contained
in the gas, as well as the coefficients of the superposition.

\end{abstract}

\maketitle
\date{\today}










Density fluctuations in dilute Bose-Einstein condensates were traditionally
described by using a mean-field theory \cite{GP61,DGPS99}, and a similar
approach has also been proposed specifically for the explanation of the
properties of elongated pencil shaped samples \cite{KNSQ00}, observed also in
experiments \cite{MSKE03,BBDH01}. Theoretical descriptions of such quasi
one-dimensional systems have used the hydrodynamic approximation
\cite{MVCT01,MS02} and have described the properties of the fluctuations as
corrections to an approximate static density. Problems, however with time
dependent mean field theories have been pointed out in \cite{GW00}. More
recent experiments reported the confinement of Rb atoms in a quantum wire
geometry \cite{PWM04,KWW04}, where the ratio of the interaction to kinetic
energy significantly exceeds unity thus approaching the Tonks-Girardeau (TG)
limit. These works have turned the purely mathematical model considered in
classic papers \cite{T36,G60,LL63} into a real physical system with potential
applications. For recent reviews see \cite{PS08,BDZ08}.

In the case of strongly interacting bosons, when the TG model can be applied
\cite{PSW00,DLO01,OD03,PS03} one can start from the many body wave function of
the system and consider the exact time dependence of the density determined by
the trapping frequency. These space and time dependent oscillations give rise
to a modulation of a weak probing field which can be observed in principle,
and can provide information on the properties of the condensate without
destroying it.

We give first a microscopic explanation of the observed oscillations based on
the theory of a strongly interacting, one dimensional, harmonically trapped
Bose gas. In this framework the ground state wave function of the system is
equivalent to a noninteracting one dimensional Fermi system \cite{G60}. Its
density oscillations are the consequence of the quantum mechanical
superposition of the ground state with energy $E_{0}$, and one or a few
excited states. Considering the simplest case of having only a single
excitation with energy $\hbar\omega$, where $\omega$ is the angular frequency
determined by the harmonic trapping potential, the wave function \ of the
condensate is%
\begin{equation}
\Psi_{\mathrm{S}}(x_{1},\ldots,x_{N},t)=c_{0}\Psi_{0}e^{-i\frac{E_{0}}{\hbar
}t}+c_{1}\Psi_{1}e^{-i(\frac{E_{0}}{\hbar}+\omega)t} \label{wf}%
\end{equation}
where $\Psi_{0}$ is the ground state and $\Psi_{1}$ is the first excited
stationary state, both being symmetric functions of the particle coordinates:
$x_{1},\ldots,x_{N}.$ This wave function yields the time dependent particle
density%
\begin{equation}
\varrho(x,t)=N%
{\textstyle\int}
|\Psi_{\mathrm{S}}(x,x_{2}\ldots,x_{N},t)|^{2}dx_{2}\ldots dx_{N}
\label{rodef}%
\end{equation}
which will obviously exhibit oscillations with frequency $\omega.$ The
explicit form of these are%
\begin{equation}
\varrho(x,t)\!=\!\left\vert c_{0}\right\vert ^{2}\!\varrho_{0}(x)+\left\vert
c_{1}\right\vert ^{2}\!\varrho_{1}(x)+2\operatorname{Re}c_{1}^{\ast}%
c_{0}\varrho_{01}(x)e^{-i\omega t} \label{rot}%
\end{equation}
where $\varrho_{0}$ and $\varrho_{1}$ are the particle densities of the ground
and excited states respectively, while $\varrho_{01}(x)=N%
{\textstyle\int}
\Psi_{0}^{\ast}(x,x_{2}\ldots,x_{N})\Psi_{1}(x,x_{2}\ldots,x_{N})dx_{2}\ldots
dx_{N}$ determines the spatial pattern of the density oscillations.

In order to calculate explicitly the terms in this sum we recall the
construction given for $\Psi_{0}$ \cite{G60,GWT01}. The ground state $N$
particle wave function can be written as%
\begin{equation}
\Psi_{0}=\frac{1}{\sqrt{N!}}\det_{(n,j)=0,1}^{(N-1,N)}\varphi_{n}(x_{j})%
{\textstyle\prod\limits_{1\leq j<k\leq N}}
\text{sign}(x_{k}-x_{j}) \label{Psi0}%
\end{equation}
where
\begin{equation}
\varphi_{n}(x_{j})=\sqrt{\frac{1}{2^{n}\,n!\ell\sqrt{\pi}}}e^{-x_{j}^{2}%
/2\ell^{2}}H_{n}\left(  x_{j}/\ell\right)
\end{equation}
is the $n$th normalized harmonic oscillator eigenfunction, where $\ell
=(\hbar/m\omega)^{1/2}$, $m$ is the mass of a particle, while the product of
the \textsl{sign} functions ensures the symmetric nature of the total wave
function. In the case of \emph{real }one-particle eigenfunctions -- as is the
case now -- instead of multiplying the determinant with the product of the
\textsl{sign} function we can take the absolute value of the determinant.
Similarly the first excited many body state $\Psi_{1}$ is obtained by
replacing the last row in the determinant in Eq. (\ref{Psi0})\ by functions of
the $N$-th excited state of the oscillator. Then, expanding according to the
last row we get
\begin{equation}
\Psi_{1}=\left\vert \frac{1}{\sqrt{N!}}%
{\textstyle\sum\limits_{j=1}^{N}}
(-1)^{N+j-1}\varphi_{N}(x_{j})D_{j}\right\vert
\end{equation}
where $D_{j}$ denotes the $(N,j)$-th minor\ of the determinant in $\Psi_{0}.$
Due to the orthogonality of the single particle functions we obtain
\begin{equation}
\varrho_{0}(x)=\!%
{\textstyle\sum\limits_{n=0}^{N-1}}
\left\vert \varphi_{n}(x)\right\vert ^{2},\quad\varrho_{1}(x)=\!%
{\textstyle\sum\limits_{n=0}^{N-2}}
\left\vert \varphi_{n}(x)\right\vert ^{2}+\left\vert \varphi_{N}(x)\right\vert
^{2} \label{ronullro1}%
\end{equation}
which can be quickly calculated using a summation formula for orthogonal
polynomials
\cite{GR07} yielding
\begin{equation}
\varrho_{0}(x)=\frac{e^{-x^{2}/\ell^{2}}}{2^{N}\ell\sqrt{\pi}(N-1)!}%
(H_{N-1}(x/\ell)H_{N}^{\prime}(x/\ell)-H_{N-1}^{\prime}(x/\ell)H_{N}(x/\ell))
\end{equation}
and similarly for $\varrho_{1}.$ The time dependent cross term contains the
product of two determinants both of which can be expanded by their last rows
and we obtain%
\begin{equation}
\Psi_{0}\Psi_{1}=\frac{1}{N!}%
{\textstyle\sum\limits_{j=1}^{N}}
(-1)^{j-1}\varphi_{N-1}(x_{j})D_{j}\times%
{\textstyle\sum\limits_{j=1}^{N}}
(-1)^{j-1}\varphi_{N}(x_{j})D_{j}%
\end{equation}
Because of the orthogonality of the single particle functions we obtain
\begin{equation}
\varrho_{01}(x)=\varphi_{N-1}(x)\varphi_{N}(x)
\end{equation}
which is exact in the one dimensional Tonks model with harmonic trapping.

\begin{figure}[h]
\includegraphics[width=6in]{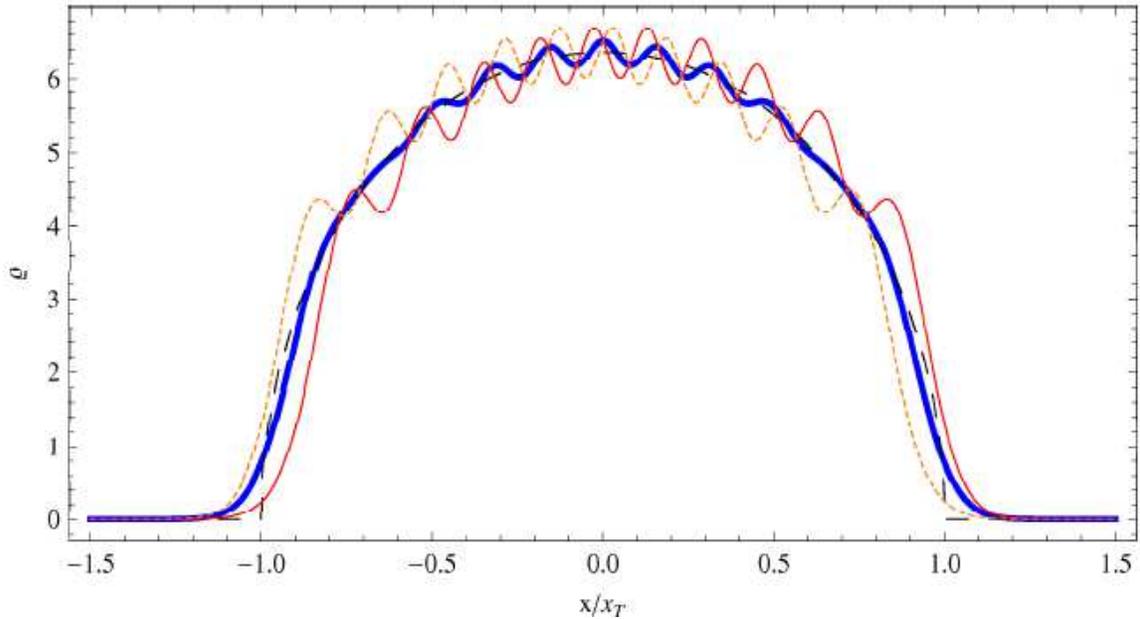}
\caption{Plot of the time
dependent particle density $\varrho(x,t)$ for 10 atoms as function of $x$
measured in units of the radius $x_{T}$. The thin solid line (red in color) is
at $t=0$, the thick solid line (blue in color) is at $t=\pi/2\omega$ and the
thin dotted line (orange in color) is at $t=\pi/\omega$. We plot the average
density $\bar{\varrho}(x)$ for comparison as a black dashed line.}%
\label{fig:rho_10atoms}%
\end{figure}

Figure \ref{fig:rho_10atoms} shows the time dependent density for 10 atoms in
comparison with the ground state average density \cite{DLO01} $\bar{\varrho
}(x)=\bar{\varrho}_{0}(1-(x/x_{T})^{2})^{1/2}$, where $\bar{\varrho}_{0}%
=\sqrt{2N}/(\pi\ell)$,\ and $x_{T}=$ $\sqrt{2N}\ell$ \ is the ground state 1D
radius of the system. $\varrho(x,t)\!$ shown here is the fundamental swinging
mode of the system, and as the number of atoms increases, it approaches the
average density, while the number of spatial oscillations increases.

We now consider the signatures of these oscillations in the time-dependent
transmission of a weak CW field $\mathcal{E}(x,t),$ which is sent through the
condensate along its axis. This field is assumed to be far-detuned from a
resonant transition of the atoms of the condensate, and will create a
polarization density along the sample. The latter can be given as
$\mathcal{P}=\mathcal{N}d^{2}\mathcal{E}/\hbar(\Delta-i\gamma)$, where
$\mathcal{N}$ is the volume density of the atoms, $d$ is the dipole matrix
element and $\gamma$ is the width of the transition in question, while
$\Delta=\omega_{0}-\omega_{L}$ is the detuning between the resonant transition
frequency $\omega_{0}$ and the carrier of the probing field $\omega_{L}$. We
write $\mathcal{N=N}_{a}\varrho(x,t)$, where $\mathcal{N}_{a}$ is the average
number of atoms in unit cross section.
In real experiments \cite{MSKE03,PWM04,KWW04} one has a lattice of pencil
shaped \ samples, therefore the actual value of $\mathcal{N}_{a}$ is the
inverse of the cross section
of one such \textquotedblleft pencil\textquotedblright. (We do not consider
here the effect of light propagating between these pencils.) The polarization,
$\mathcal{P}$ leads to a space and time dependent susceptibility and index of
refraction
\begin{equation}
n(x,t)=\left(  1+\frac{\mathcal{N}_{a}d^{2}}{\epsilon_{0}\hbar(\Delta
-i\gamma)}\varrho(x,t)\right)  ^{1/2} \label{ir}%
\end{equation}
As the \ time dependence of the harmonic trap is very slow with respect to the
frequency of the optical fields, instead of the wave equation we shall solve
the one dimensional amplitude equation for the electric field:
\begin{equation}
\frac{\partial^{2}}{\partial x^{2}}E(x,t)+n^{2}(x,t)k^{2}E(x,t)=0
\label{ampeq}%
\end{equation}
where $k=2\pi/\lambda_{0}$ is the corresponding wave number in vacuum. Here
$E(x,t)$ is the temporally slowly varying
amplitude of the full electric field: $\mathcal{E}(x,t)=E(x,t)e^{-i\omega_{L}t}$. The solution of this
equation with a given incoming plane wave from the negative $x$ direction
shall yield the transmitted wave $E_{\mathrm{tr}}(t)e^{ikx}$ for $x\gg x_{T},$
as well as a reflected wave at $x\ll-x_{T}$. The results of a numerical
solution will be discussed below. In order to get a better insight into the
nature of the problem we also present \ an approximate analytic solution to
(\ref{ampeq}) by a 2nd order WKB\ approximation obtained for the forward
propagating wave as%
\begin{equation}
E_{\mathrm{f}}(x,t)=\frac{1}{\sqrt{n(x,t)}}E_{0}\exp\left(  ik%
{\textstyle\int\limits_{x_{0}}^{x}}
n(x^{\prime},t)dx^{\prime}-i\frac{n^{\prime}(x)}{4kn^{2}(x)}-i\int_{x_{0}}%
^{x}\frac{(n^{\prime}(x^{\prime}))^{2}}{8kn^{3}(x^{\prime})}dx^{\prime
}\right)  , \label{eq:elfieldfwd}%
\end{equation}
where $E_{0}$ is the incident field amplitude at $x_{0}$, far before the
condensate. The transmission coefficient of the condensate depends only on
time for $x\gg x_{T}$, i.e. far beyond the condensate:
\begin{equation}
T(t)=\left\vert \frac{E_{\mathrm{f}}(x,t)}{E_{0}}\right\vert ^{2}=\exp\left(
-2k\int_{x_{0}}^{x}\operatorname{Im}n(x^{\prime},t)dx^{\prime}+\frac{1}%
{4k}\int_{x_{0}}^{x}\operatorname{Im}\frac{(n^{\prime}(x^{\prime}))^{2}}%
{n^{3}(x^{\prime})}dx^{\prime}\right)
\end{equation}
since $n(x,t)=1$ for $x\gg x_{T}$.
In case of an off-resonant external field we can expand the index of
refraction given by (\ref{ir}) under the integral up to second order as
\begin{equation}
n(x,t)=1+\beta\varrho(x,t)/2-\beta^{2}\varrho^{2}(x,t)/8
\end{equation}
with $\beta=\mathcal{N}_{a}d^{2}/(\epsilon_{0}\hbar(\Delta-i\gamma))$.
In the following we consider the light source at $x_{0}=-\infty$, and a
photodetector at $x=\infty$. Substituting the second order expansion of the
refractive index into the formula for the transmission we have:%
\begin{equation}
T(t)=\exp\left(  -Nk\operatorname{Im}\beta+\frac{k\operatorname{Im}\beta^{2}%
}{4}\int_{-\infty}^{\infty}\varrho^{2}(x,t)dx+\frac{\operatorname{Im}\beta
^{2}}{16k}\int_{-\infty}^{\infty}(\varrho^{\prime}(x,t))^{2}dx\right)
\label{eq:tr1}%
\end{equation}
We write $\varrho(x,t)=r_{0}(x)+r_{1}(x)\cos(\omega t+\alpha)$, where
$r_{0}(x)=|c_{0}|^{2}\varrho_{0}(x)+|c_{1}|^{2}\varrho_{1}(x)$ is an even
function and $r_{1}(x)=2|c_{0}||c_{1}|\varrho_{01}(x)$ is an odd function of
$x$, and $\alpha$ denotes the relative phase of $c_{0}$ and $c_{1}$.
Substituting this into (\ref{eq:tr1}) and using the parity of $r_{0}$ and
$r_{1}$, we obtain the following formula for the transmission:
\begin{equation}
T(t)=T_{N}\exp\left[  \zeta\cos\left(  2(\omega t+\alpha)\right)  \right]  ,
\label{eq:tr2}%
\end{equation}
where
\begin{equation}
T_{N}=\exp\left(  -Nk\operatorname{Im}\beta+\frac{k}{4}\operatorname{Im}%
(\beta^{2})\int_{-\infty}^{\infty}\left(  r_{0}^{2}(x)+\frac{1}{2}r_{1}%
^{2}(x)\right)  dx+\frac{\operatorname{Im}\beta^{2}}{16k}\int_{-\infty
}^{\infty}\left(  (r_{0}^{\prime2}+\frac{1}{2}(r_{1}^{\prime2}\right)
dx\right) \nonumber
\end{equation}
does not depend on time, $\alpha=\arg(c_{0}-c_{1})$, and
\begin{equation}
\zeta=(2|c_{0}||c_{1}|)^{2}\operatorname{Im}(\beta^{2})(kR_{1}(N)/8+R_{2}%
(N)/32k) \label{eq:zetadef}%
\end{equation}
with the integrals
\[
R_{1}(N)=\int_{-\infty}^{\infty}\varrho_{01}^{2}(x)dx,\ \ R_{2}(N)=\int
_{-\infty}^{\infty}(\varrho_{01}^{\prime2}(x))dx
\]
depending only on the number of particles in the sample. $R_{1}$ and $R_{2}$
can be easily calculated up to several hundreds of atoms with a finite sum
expression, which can be derived using known formulas for products of Hermite
polynomials \cite{GR07}.

We can expand the time dependent factor in (\ref{eq:tr2}) into a Jacobi form
\cite{AS65}, which is directly related to the discrete Fourier transform of
the time dependent transmission:
\begin{equation}
T(t)=T_{N}\left[  I_{0}(\zeta)+2\sum_{s=1}^{\infty}I_{s}(\zeta) \cos\left(  2
s (\omega t+\alpha) \right)  \right]
\end{equation}
where $I_{s}(\zeta)$ are the modified Bessel functions. This means that the
complex Fourier coefficients $\tilde{T}$ of the time-dependent transmisson are
proportional to modified Bessel functions with the same argument,
$s=0,1,2,...$ :%
\begin{equation}
\tilde{T}(2s\omega)\sim I_{s}(\zeta)e^{-2si\alpha}.
\end{equation}
$I_{1}(\zeta)$ is real, therefore the phase of $\tilde{T}(2\omega)$ yields the
relative phase of the states $\Psi_{0}$ and $\Psi_{1}$: $\alpha=-\frac{1}%
{2}\arg\tilde{T}(2\omega)$, in the case of a positive detuning. ($I_{1}$ is
odd for real arguments, and the sign of $\zeta$ is the sign of $\Delta$.) A
well known relation for the Bessel functions \cite{AS65} enables us to
calculate $\zeta$ from the transmission spectrum:%
\begin{equation}
\zeta=\frac{2I_{1}(\zeta)}{I_{0}(\zeta)-I_{2}(\zeta)}=\frac{2\tilde{T}%
(2\omega)e^{2i\alpha}}{\tilde{T}(0)-|\tilde{T}(4\omega)|} \label{eq:zeta}%
\end{equation}

Since $\zeta$ is composed of quantities which characterize the condensate and
its interaction with the CW field, the knowledge of $\zeta$ gives information
about these quantities. E.g. if we know everything in $\zeta$ except for the
coefficients $c_{0}$ and $c_{1}$, then a measurement of the time-dependent
transmission yields the value of $|c_{0}||c_{1}|$ and the relative phase
$\alpha$ which (using $|c_{0}|^{2}+|c_{1}|^{2}=1$) allows us to calculate
$c_{0}$ and $c_{1}$ up to a global phase factor.
Alternatively, the number of particles in the condensate can be calculated
from the Fourier coefficients of the time-dependent transmission, if the other
quantities in $\zeta$ are already known. This means that one can accurately
measure the number of particles in a condensate without destroying it.

\begin{figure}[ptb]
\includegraphics[width=3in]{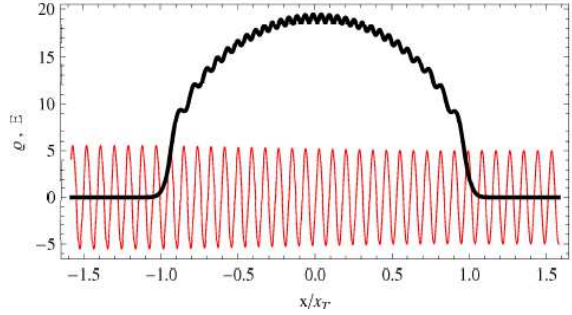}
\caption{Plot of the electric field amplitude $E(x/x_{T},t=0)$, resulting from
the numerical solution of Eq. (\ref{ampeq}), with thin solid line (red in
color), in arbitrary units, as a probing field passing through the condensate of 30 atoms
with particle density $\varrho(x/x_{T},t=0)$, plotted as a thick black line.
The approximate analytic solution given by Eq.(\ref{eq:elfieldfwd}) differs
from the numerical result by less then the line thickness in this figure.}%
\label{fig:transmittedfield}
\end{figure}

\begin{figure}[ptb]
\includegraphics[width=6in]{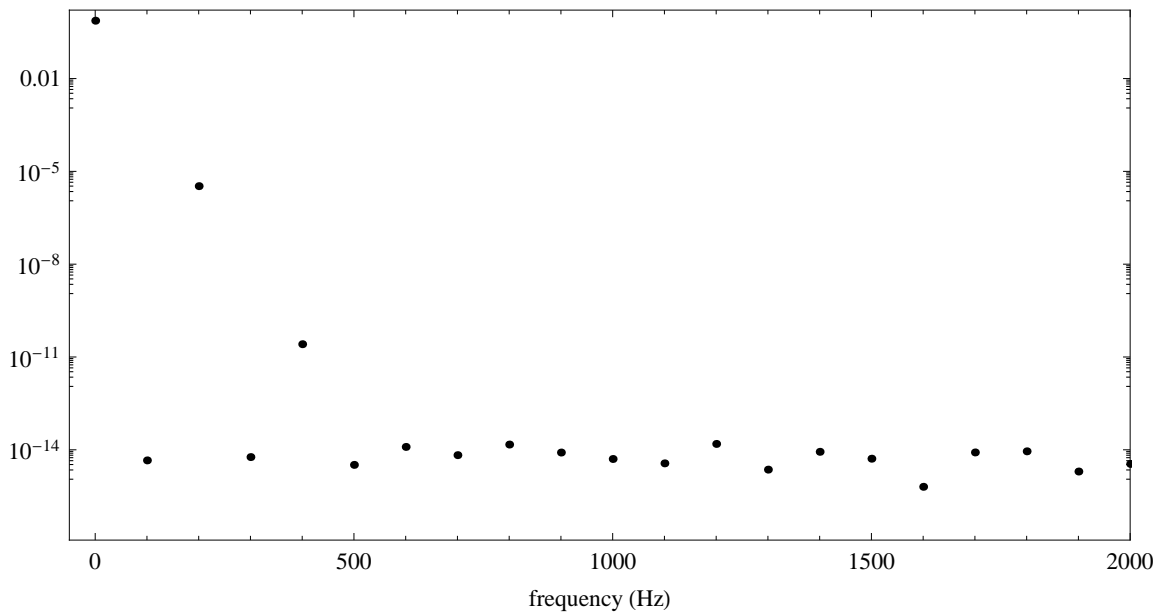}\caption{Plot
of the magnitude of the Fourier coefficients of the time dependent
transmission.}%
\label{fig:transspectrum}%
\end{figure}

We illustrate the use of the modulated transmission for this latter case, by
processing a simulated transmisson signal which we obtain from the numerical
solution of the second order amplitude equation (\ref{ampeq}). We assume a
sample where $^{87}$Rb atoms are trapped in an array of pencil shaped
condensates \cite{MSKE03,PWM04,KWW04} containing 30 atoms in a superposition
state (\ref{wf}) with $c_{0}=c_{1}=1/\sqrt{2}$. The laser light for the
transmission measurement is assumed to be detuned with an angular frequency
$\Delta=2\pi\times90$ MHz from the center of the $D_{1}$ line ($\lambda_{0}%
=$794.978 nm), and we use $\gamma=1.80647\times10^{7}$ 1/s and $d=1.4651\times
10^{-29}$ Cm \cite{RbData}. Fig. \ref{fig:transmittedfield} shows the density
$\varrho(x,t=0)$ and the electric field amplidude $E(x,t=0)$ for 30 atoms
obtained from the solution of Eq. (\ref{ampeq}).
Fig. \ref{fig:transspectrum} shows the Fourier amplitudes of the
time-dependent transmission, assuming a longitudinal trap angular frequency
$\omega=2\pi\times100$ Hz. A calculation based on the data shown in this
figure and using Eqs. (\ref{eq:zetadef}) and (\ref{eq:zeta}) reproduce
correctly that the sample contains 30 or 31 atoms. 

In conclusion, we have constructed an exact many body superposition state for
the Tonks gas, exhibiting a time-dependent particle density in a swinging
mode. Generalizations for superpositions involving higher excited modes are
straightforward. The model for the interaction of the Tonks gas with a weak
laser beam opens the possibility of measuring the effect of these density
oscillations as a weak but measurable oscillation in the transmission signal.
The approximate analytic formula obtained for this time-dependent transmission
allows for the calculation of the quantities which characterize the condensate
and its interaction with the CW field:  the coefficients of the many body
superposition state, or the number of atoms in the Tonks gas could be measured
without destroying the sample.

This work was supported by the Hungarian Scientific Research Fund OTKA under
Contracts No. T48888, M36803, M045596. We thank P. Földi for useful discussions.

\end{document}